# Parking Space Management via Dynamic Performance-Based Pricing

Daniel Mackowski, Yun Bai, Yanfeng Ouyang[1]

Department of Civil and Environmental Engineering, University of Illinois at Urbana-Champaign, Urbana, IL 61801, USA

**Abstract**
In congested urban areas, it remains a pressing challenge to reduce unnecessary vehicle circling for parking while at the same time maximize parking space utilization. In observance of new information technologies that have become readily accessible to drivers and parking agencies, we develop a dynamic non-cooperative bi-level model (i.e. Stackelberg leader-follower game) to set parking prices in real-time for effective parking access and space utilization. The model is expected to fit into an integrated parking pricing and management system, where parking reservations and transactions are facilitated by sensing and informatics infrastructures, that ensures the availability of convenient spaces at equilibrium market prices. It is shown with numerical examples that the proposed dynamic parking pricing model has the potential to virtually eliminate vehicle circling for parking, which results in significant reduction in adverse socioeconomic externalities such as traffic congestion and emissions.

## 1. Introduction

In practically all major cities, the time and frustration associated with finding a parking space during peak hours not only upsets drivers, but also significantly decreases the city's economic, environmental, and social sustainability. The economic impact, measured in terms of wasted resources (time or fuel) and lost economic potential, is the most visible. On average, 8.1 minutes are wasted each time a driver circles around a U.S. city in search of a parking space (Shoup, 2005). This extra circling accounts for up to 30% of urban traffic congestion, and generates an extra 4,927 vehicle-miles-travelled (VMT) per parking space per year (Shoup, 2006). A large portion of this circling occurs in densely developed downtown areas. Additionally, according to a survey in Year 2011, 60% of the responded drivers reported that at least once they were so frustrated searching for parking that they eventually gave up, leaving behind only congestion and lost economic opportunity (IBM, 2011). The environmental impacts are also significant. In a large city such as Chicago (with over 35,000 parking spaces), the excess vehicle distance caused by circling burns 8.37 million gallons of gasoline into an extra 129,000 tons of $CO_2$ each year (Ayala et al., 2012).

Due to these economic and environmental issues, management of urban parking, via pricing schemes for example, is often necessary. Such efforts can impose significant social impacts, including direct influences on travel demand and equity issues. For example, outdated parking policies such as free or tax-exempt parking in essence provide a subsidy to those owning and operating a vehicle, which may be worth hundreds of dollars per motorist per year (Litman, 2005). This subsidy, often also seen as one approach to demand management, tends to favor wealthier car-owning households and may divert investments from alternative transportation modes (such

---

as pedestrian, bicycling, or transit). However, regardless of the city's mode split, it is important to efficiently utilize the parking resources allocated to car owners, as they are also important customers of the local economy. While easy to implement, spatially and/or temporally invariant parking management strategies, common in many cities' parking meter programs, may not be able to optimally regulate parking demand, which results in congestion and lost economic potential.

The rapid advances of information technology in recent years hold the promise to bring in a new paradigm of parking management systems that can reduce the negative externalities associated with urban parking. However, this new paradigm, commonly referred to as "smart parking," faces three types of barriers: political, technological, and operational. The political and technological barriers were lowered significantly over the past decade with the empirical implementation of several full-scale systems that include real-time occupancy sensing, public information dissemination, electronic payments, and so forth. The pilot SF*park* variable pricing program in San Francisco, for example, has shown the possibility of overcoming the political barrier to time-varying parking policies. Hence, the biggest remaining barrier is operational, which focuses on effectively managing urban parking while considering infrastructures, policies, enforcement, and general operations. In particular, demand management, often accomplished through parking pricing, is critical to efficient parking operations. To the authors' best knowledge, no demand management strategies yet exist (beyond experimental) that are able to utilize dynamic and real-time information to improve efficiency, equity, and the user experience and in turn mitigate congestion and other social and environmental issues. The working mechanism of such a management system should also be easily understood by drivers and other stakeholders.

Even with real-time information, the implementation of new parking systems that will affect thousands of people every day is complex. Most existing models predict travel and parking demand based on historic data, but unfortunately fail to address complex interactions among various parking supply/demand factors and the interrelationships among pricing/travel/parking decisions. In reality, the decision-making process of the stakeholders (e.g. parking agency, individual drivers) often involves complicated gaming behaviors. On one hand, the agency's parking management strategies (e.g. pricing) directly affect drivers' travel and parking decisions, which in turn influence the parking demand. On the other hand, drivers with different origins, destinations, and socioeconomic characteristics compete against one another for limited parking resources at popular locations, and this competition eventually shapes the parking market and pricing at equilibrium. Integrating multiple layers of such decisions into one overarching modeling framework is challenging, as it involves different stakeholders who have independent and sometimes conflicting objectives.

This paper presents a new parking pricing and management system that incorporates ideas from variable pricing (e.g. SF*park*), on- and off-street parking reservation systems (e.g. Xerox, ParkWhiz), game theory, and downtown parking economics (see Arnott et al., 2013). The primary goal of this paper is to develop a dynamic, demand based real-time pricing model to optimally allocate parking spaces in busy urban centers, thus reducing congestion and other negative economic and environmental costs. This pricing model is online in nature and is able to react to real-time demand variations. It allows a parking agency to set system optimal pricing policies (e.g. to minimize congestion, maximize social surplus, maximize revenue) while considering user competition and market equilibrium. Additionally, analysis of this pricing model provides insights into practical solution techniques for complex bi-level programming problems, i.e. mathematical programs with equilibrium constraints (MPECs), in dynamic settings.

The remainder of this paper is organized as follows. Section 2 reviews the literature related to smart parking, the existing variable pricing systems, and state-of-the-art modeling methods. Section 3 presents the assumptions, notation, and formulation of the proposed non-cooperative Stackelberg leader-follower game model for dynamic parking pricing. A solution approach is developed by transforming the bi-level MPEC into a solvable single-level mixed-integer quadratic program (MIQP) that is implemented in a rolling-horizon scheme. Section 4 illustrates the performance and potential impacts of the proposed dynamic pricing model using a numerical example based on an urban neighborhood in SF*park*'s current system. Section 5 concludes the paper and discusses possible future research.

## 2. Literature Review

Problems associated with urban parking have gained considerable attention over the past decade. Shoup (2005) provides a meticulous summary of the status-quo of parking management and a number of innovative ideas. Using ideas from the book, the pilot SF*park* program successfully applied demand responsive performance-based pricing to seven neighborhoods in San Francisco. The underlying theory is that if there is always one open space on each block, drivers can immediately find an open space that suits their preferences, which would virtually eliminate circling for parking. The program uses recorded occupancy data to raise or lower meter and garage prices every six weeks with the goal of achieving a target occupancy level (e.g. 85% on average) on every street block during every time period (e.g. morning, afternoon, evening). Another pricing model by Xerox® utilizes historic data to predict future parking demand and set parking prices accordingly. This model, currently implemented in Los Angeles's LA ExpressPark™ program, can take special events into account, but also updates pricing tables rather infrequently (i.e. only once every several weeks). Unfortunately, for both programs, such long intervals between price updates limit the models' ability to handle anything other than an average parking demand.

It is now possible to collect accurate information dynamically and use it to predict real-time demand (Caicedo et al., 2012). Information technology and its subset of "smart parking" applications have evolved rapidly over the past decade. Several companies have heavily invested in "smart parking" technologies and are continuously improving their products as competition heats up to serve the growing demand (e.g. Streetline, Fybr, Deteq, Libelium, etc.). Current state-of-the-art parking technologies include hockey puck sized sensors recording real-time occupancy data, instant electronic payment options, dynamic message signs, smartphone applications that allow reservation of garage parking, and much more. Communication between the parking agency and users is typically via smartphone and web applications, with forthcoming integration into vehicles' on-board navigation systems. Despite all these technological advances, however, few existing urban parking models are currently capable of utilizing such real-time information to enhance urban parking.

The transition to a more responsive and dynamic model is not trivial, however. Current efforts are limited to the closed confines of parking garages assuming that the operator has complete control of the environment, which is rarely the case in reality. For example, Geng and Cassandras (2011) developed a system to dynamically assign and reserve user optimal parking spaces in real-time for a single parking garage. Such model assumes a centralized system that unfortunately does not capture drivers' independent/competing decision-making processes and the agency's pricing leverage. Nor does it consider congestion reduction as an important objective for on-street parking. Ayala et al. (2011; 2012) developed models that help authorities set parking prices that attempt to minimize system-wide driving distances. Their pricing schemes are offline in nature, assuming a static parking demand that does not change over time. Therefore, it is not clear how their models can be implemented in a large-scale and dynamic real-world environment.

Recently, Qian and Rajagopala (2013) developed a model showing the promise of real-time pricing in achieving system optimal objectives. Given the dynamic traffic demand on a general network, and assuming user equilibrium travel behavior, they show that lot-based pricing schemes can achieve system optimal parking flow patterns that minimize total cost, including users' parking search time within the parking facility. Further, because parking price and parking occupancy assume a one-to-one mapping in their model, optimal real-time prices can be set solely based on real-time occupancy levels, provided the information is disseminated to users in real-time. This would allow parking pricing changes similar to SF*park* except in real-time, if their assumptions regarding approximations to parking search times prove valid. However, the typical search time function used as the basis in their model was developed for a German off-street parking guidance system and may not apply directly to on-street search times, although results from Levy et al. (2012) suggest a similar shape. Regardless, the paper focused on morning commute scenarios for large centrally owned off-street parking facilities. Further research into all-day on- and off-street parking scenarios should provide interesting results appropriate for downtown urban areas posed with challenging parking management situations. The management of such parking systems involves non-cooperative decision making between the parking agency and the users in the network. However, in their paper, the dynamic pricing from the parking agency's point of view is not discussed, including the impact of travelers' choices on the agency's optimal strategy accounting for revenue and/or other social objectives.

Game theoretic models that allow hierarchical yet integrated decision making are particularly useful for this parking management problem. The Stackelberg leader-follower model and its MPEC formulations have been widely applied to hierarchical decision making in transportation planning, economics, engineering design, etc. Luo et al. (1996) provided a good review of the fundamental results in MPECs. Such models effectively capture multiple stakeholders' individual decisions while designing management strategies to achieve a system objective -- such as the transportation network design (NDP) problems in Abdulaal and Leblanc (1979), Yang and Bell (2001), Small, (1992), Giuliano, (1994), Daniel, (1995), Lu (2006), etc.; and competitive supply chain problems in Bai et al. (2012, 2014). MPECs are in general difficult due to the inherent non-convexity and non-smoothness. Solution methods were developed for the linear/quadratic program with linear complementarity constraints (LP/QPCC), e.g. see Bai et al. (2013) and Hu et al. (2008; 2012), and discretely constrained (DC-) MPECs (Bai et al., 2012; 2014). However, the problem scales up rather quickly such that the existing methods' ability to solve larger problem instances is very limited. In this paper, we develop a dynamic MPEC model for one parking agency and many independent users, with the objective of finding optimal prices and equilibrium parking choices for an urban neighborhood in a multi-interval time horizon. Due to high complexities, we develop a rolling horizon solution approach that is able to integrate both historic data and real-time information into the optimal solutions for each time interval. The static MPEC problem for each time interval is then reformulated into an equivalent quadratic MIQP that can be solved by existing solvers.

## 3. Methodology

Advancing the parking management paradigm to a dynamic and performance-based one requires the development of a reliable pricing model as well as the design of the technological infrastructure within which the model will operate. Section 3.1 below briefly introduces the technological infrastructure requirements. Section 3.2 outlines the model formulation and associated assumptions for the pricing model, while Section 3.3 presents the solution techniques.

### 3.1. System Infrastructures

Mackowski et al. (2014) provides an overview of the infrastructure requirements and implementation details. Most of the technological infrastructure can be built upon that of the current "smart parking" systems; e.g. those providing space availability information as well as route guidance (Parker™ by Streetline, Merge® by Xerox, etc.) to the users in real-time. Nevertheless, several key components must be added to allow the pricing model to perform to its full potential.

The primary component needed is a reservation system for on-street parking. Currently, parking garages are able to offer reservations to users via a website or smartphone app (e.g. SpotHero, ParkWhiz) to ensure that a space is available when the user arrives. Extending this concept to on-street parking is possible if access to spaces can be similarly controlled. Networked indicators that light up red when a space is reserved are inexpensive and easy to install on parking meters. To honor parking reservations, parking enforcement officers may have access to a communication system that sends them the correct license plate and alerts them when and where a vehicle is parked illegally so they can target their enforcement and lower the rate of violations. PARC (Xerox's research group) is currently developing prototypes of this technology to demonstrate its feasibility in real-world urban applications (U.S. Patent No. 8,671,002, 2014). This reservation component will expand the market for available parking spaces (from just local vehicles to all users travelling to nearby destinations), allow users to quickly find the best available space during each trip, and provide better user experiences. Additionally, the recent work by Liu et al. (2013) has shown that devoting a portion of the overall parking supply to spaces requiring reservations can reduce bottleneck congestion by smoothing out traffic arrivals – at least during the morning commute.

The technological infrastructure must also collect live data to support pricing decisions in real-time. Partial demand data will be provided by the reservation system, and complete occupancy information will be provided by sensors in each parking space. Such data streams, after being cleaned of any noise or inaccuracies, will be combined with historic occupancy and demand data, as well as other relevant data (e.g. street closures, sports events), to provide adequate information for real-time future demand prediction. These real-time data serve as the input for the proposed model to make pricing adjustments.

### 3.2. Model Formulation

Now we will present a dynamic MPEC formulation of a Stackelberg leader-follower game model that determines parking prices and management strategies in real-time. This model addresses spatial and temporal variations in parking demand, as well as complex interactions among multiple stakeholders.

We consider a city neighborhood and a set of discrete time intervals in a time horizon $T$. A set of discrete parking areas $J$ within the neighborhood is defined by block face for on-street parking and/or by parking lot for off-street parking. $J$ includes only the parking areas in our system within the neighborhood, and other nearby parking areas, such as private parking or areas managed by other agencies, are termed "outside lots."[2] Area $j \in J$ has a capacity $c_j$. The parking agency's main decision is the price of each parking area at each time interval $\mathbf{p} \triangleq \{p_j^t\}$. The traffic demand is generated from a set of origins $O$, e.g. traffic analysis zones (TAZs), and goes toward a set of destinations $D$. A driver from origin $o \in O$ to destination $d \in D$ may arrive at the parking neighborhood at time $t \in T$ and have a desired parking duration of $n \in \{1, 2, ..., N\}$ time intervals. The total elastic demand of each type of driver (i.e. whom we will call "users"), specified by the origin, destination, arrival time, and parking duration, follows an inverse demand function $H(\cdot)$ of the equilibrium disutility $u_{od}^{t,n}$ of this type of demand. The equilibrium disutility $u_{od}^{t,n}$, identical for all users of this type regardless of their choice of parking area $j$, is measured by the parking price $p_j^t$, the cost for required driving $v_{oj}$ between origin $o$ and area $j$, and that for walking $w_{jd}$ between area $j$ and destination $d$. The costs for the latter two can be estimated from the respective indifference curves (i.e. $\theta$ and $\theta'$) between monetary cost and travel times, assuming perfect substitution. We let $h_{j,od}^{t,n}$ be the number of users choosing to park at area $j$. The limited total parking capacity across all areas in our system may not be sufficient or competitive to serve all the demand. In such cases, part of the users are assumed to lose service or choose lots outside of our system. For model simplicity, we do not consider the detailed spatial distribution of outside lots, but rather include a dummy lot, denoted by $\zeta$ with sufficiently large capacity to absorb all excessive demand that is lost from our system. Hence, $\sum_{j \in J \cup \{\zeta\}} h_{j,od}^{t,n} = H(u_{od}^{t,n})$. Finally, it is assumed that users lock in and pay the same price $p_j^t$ per time interval if they start to park at interval $t$. The occupancy of parking area $j$ at time $t$, $f_j^t$, satisfies conservation over time based on the amounts of entering users, $\sum_{n \in \{1,2,...N\}} q_j^{t,n}$, and departing users, $g_j^t = \sum_{m=\max(1,t-N)}^{t-1} q_j^{m,t-m}$.

The mathematical model, developed from the parking agency's perspective, takes the form of a dynamic MPEC. It has the parking agency's problem in the upper level and users' collective parking choice equilibrium in the lower level. The parking agency is the leader and sets the time- and location-specific prices. The primary objective is to minimize the absolute difference of actual occupied parking spots in each parking area $j$ from its target level $\kappa_j c_j$, where $\kappa_j$ is the ideal percentage occupancy, at a penalty of $\beta$ per spot per time interval. Secondary objectives such as maximizing social surplus (consumer plus agency surpluses) and maximizing revenue can also be added with respective weights $\alpha_s$ and $\alpha_r$. This is illustrated in the objective function of the upper level problem below.

*(upper level)*

$$\min_{\mathbf{f,p,q,h,u,g}} \beta \sum_{t=t_s}^{t_s+\tau} \sum_{j \in J} \left| \kappa_j c_j - f_j^t \right| - \alpha_s \sum_{t=t_s}^{t_s+\tau} \sum_{j \in J} \left( \sum_{n \in \{1,2,...N\}} n \cdot p_j^t \cdot q_j^{t,n} + \sum_{n \in \{1,2,...N\}} \sum_{o \in O} \sum_{d \in D} \frac{1}{2} h_{j,od}^{t,n} (u_{\max} + u) \right) \quad (1.1)$$
$$- \alpha_r \sum_{t=t_s}^{t_s+\tau} \sum_{j \in J} \sum_{n \in \{1,2,...N\}} n \cdot p_j^t \cdot q_j^{t,n}$$

---

[2] Private parking areas outside the agency's ownership could be included within our system if the parking prices and availability, as well as the capacity and location, are known at each time interval.

$$\text{s.t. } -\varepsilon_l \leq p_j^t - p_j^{t-1} \leq \varepsilon_r, \quad \forall j \in J, t \in \{t_s, ...t_s + \tau\} \tag{1.2}$$

$$\gamma_j \leq p_j^t \leq \eta_j, \quad \forall j \in J, t \in \{t_s, ...t_s + \tau\} \tag{1.3}$$

$$f_j^t = f_j^{t-1} - g_j^t + \sum_{n \in \{1,2,...N\}} q_j^{t,n}, \quad \forall j \in J, t \in \{t_s, ...t_s + \tau\} \tag{1.4}$$

$$q_j^{t,n} = \sum_{o \in O} \sum_{d \in D} h_{j,od}^{t,n}, \quad \forall j \in J, n \in \{1,2,...N\}, t \in \{t_s, ...t_s + \tau\} \tag{1.5}$$

$$g_j^t = \sum_{m=\max(1,t-N)}^{t-1} q_j^{m,t-m}, \quad \forall j \in J, t \in \{t_s, ...t_s + \tau\} \tag{1.6}$$

Note that the model is formulated on a rolling horizon of $\tau$ intervals, for all $t_s = 1, 2, ..., T$. The first term in (1.1) pushes parking space allocation toward the target occupancy levels. Social surplus is captured by the second term in (1.1). It only considers direct consumer and agency surpluses and does not include external benefits or impacts such as congestion, vehicular emissions, and so on. The third term captures the agency's parking pricing revenue. The most likely constraints that the agency must satisfy are those related to pricing policies set by city legislation or those related to public acceptance. In the above formulation, Constraints (1.2) and (1.3) impose an upper $\varepsilon_r$ and lower $\varepsilon_l$ limit to price variations between consecutive time intervals, and a minimum $\gamma_j$ and maximum $\eta_j$ price that each parking area may take, respectively.[3] Constraints (1.4) preserve the conservation of available parking spaces between time intervals. Constraints (1.5) and (1.6) define the total demand arriving at and leaving each parking area at time interval $t$, respectively.

 A measure of system performance could be the amount of demand that choose not to use our system (i.e. defined as "lost users"). Some users will inevitably be "lost" to parking outside of the system (e.g. hotel, restaurant).[4] However, users can also be lost when they cannot find an available parking space and give up searching, either taking their business elsewhere or going home, which is undesirable. Therefore, this model sets prices near market prices[5] to encourage as many users as possible to use this system while allocating them optimally and without exceeding the target occupancy level. Additional performance metrics will be discussed below in the context of the model's lower level formulation.

 In the lower level, individual drivers act as followers. They decide where to park based on the spatial distribution of available parking spaces, prices, and their travel origins/destinations. In a busy urban neighborhood, we assume that numerous users make decisions simultaneously in each time interval, and hence their problem resembles a type of Nash equilibrium, i.e. traffic network equilibrium; see Sheffi (1985) for a detailed review.

 To derive the lower level equilibrium condition, we first formulate the decision problem of each individual user. To this end, we define $L_{od}^{t,n}$ as the set of drivers who travel between the origin-destination pair $od \in OD$ at time interval $t$ who have parking duration $n$, and $\left|L_{od}^{t,n}\right| = \sum_{j \in J \cup \{\zeta\}} h_{j,od}^{t,n}$.

---

[3] It is also possible to fix garage rates during certain times-of-day, or offer early bird and other discounts, both of which can be easily incorporated into the model through additional constraints.

[4] Even if parking is free however, there is still some disutility related to driving to a parking space and walking to the destination. It is assumed that there is a market price for parking, public and private, within the area. Therefore, any user that does not use the system due to excess disutility (as opposed to trouble finding a parking space) is assumed to use some alternative parking space in the neighborhood, which is likely a private parking lot such as a restaurant or hotel that can provide customers free or discounted pricing.

[5] Because of the policy constraints (1.2)-(1.3), prices may not be at exact market equilibrium.

We then define binary variable $\mathbf{z}_l = \{z_{jl}, \forall j \in J \cup \{\zeta\}\} \in \{0,1\}^{|J|}$ as the parking decision of driver $l \in L_{od}^{t,n}$, i.e. driver $l$ chooses to park in area $j$ if $z_{jl} = 1$; or $z_{jl} = 0$ otherwise. We further define disutility $\phi_{j,od}^{t,n} = n \cdot p_j^t + \theta w_{jd} + \theta' v_{oj}$ for users from $od \in OD$ to park in area $j \in J$ for duration $n \in \{1,2,...,N\}$ at time interval $t \in T$. The disutility of the dummy lot $\phi_{\zeta,od}^{n,t} = \lambda_{\zeta,od}^{n,t}$ is defined as

$$\lambda_{\zeta,od}^{t,n} = \min\left\{H^{-1}(0),\ \max_{j \in J}\left\{n\left[p_j^{t-1} + (t+1-t_s)\varepsilon_r\right] + \theta w_{jd} + \theta' v_{oj}\right\}\right\}, \tag{2}$$

$\forall o \in O, d \in D, n \in \{1,2,...N\}, t \in \{t_s,...t_s + \tau\}$.

This is set high enough so that users are assigned to the dummy lot only when real lots are all full, but no higher than the intercepts of the demand curves.

Then for each user $l \in L_{od}^{t,n}$, the parking decision is made based on the following disutility minimization problem:

$$\min_{\mathbf{z}_l} \sum_{j \in J \cup \{\zeta\}} \phi_{j,od}^{t,n} \cdot z_{jl} \tag{3.1}$$

s.t. $z_{jl} \in \{0,1\}, \forall j \in J \cup \{\zeta\}$, (3.2)

$$\sum_{j \in J \cup \{\zeta\}} z_{jl} = 1, \forall l, \tag{3.3}$$

$$z_{jl} + \sum_{-l \in L_{od}^{t,n}, \forall od \in OD, n \in \{1,2,...,N\}} z_{j,-l} + f_j^{t-1} - g_j^t \leq c_j, \forall j \in J. \tag{3.4}$$

Constraints (3.2) and (3.3) make sure that only one parking lot is selected by any user. Constraints (3.4) are lot capacity constraints coupled among all users in $L_{od}^{t,n}$. It can be seen that problem (3.1)-(3.4) is a totally unimodular assignment problem, and thus the binary constraint (3.2) can be relaxed to

$$0 \leq z_{jl} \leq 1, \forall j \in J \cup \{\zeta\}. \tag{3.5}$$

We further define $\boldsymbol{\lambda}_l = \{\lambda_l\}, \boldsymbol{\rho}^t = \{\rho_j^t\}, \boldsymbol{\mu}_l = \{\mu_{jl}\}$ as dual variables corresponding to Constraints (3.3), (3.4) and (3.5), respectively. Therefore, we are able to further derive the Lagrangian of problem (3.1) and (3.3)-(3.5) as follows:

$$\mathcal{L}(\boldsymbol{\mu}_l, \boldsymbol{\lambda}_l, \boldsymbol{\rho}^t) = \min_{\mathbf{z}_l \geq \mathbf{0}} \sum_{j \in J \cup \{\zeta\}} \phi_{j,od}^{t,n} \cdot z_{jl} + \sum_{j \in J \cup \{\zeta\}} \mu_{jl} \cdot (z_{jl} - 1) - \lambda_l \cdot \left(\sum_{j \in J \cup \{\zeta\}} z_{jl} - 1\right) + \\ \sum_{j \in J} \rho_j^t \left(z_{jl} + \sum_{-l \in L_{od}^{t,n}, od \in OD, n \in \{1,2,...,N\}} z_{j,-l} + f_j^{t-1} - g_j^t - c_j\right). \tag{4}$$

For the ease of notation, we omit the sub/superscripts $t$, $n$, and $od$ in the remainder of Section 3. As such, we derive the first order Karush-Kuhn-Tucker (KKT) conditions for each individual user's problem as follows:

$$0 \leq z_{jl} \perp \phi_j + \mu_{jl} - \lambda_l + \rho_j^t \geq 0, \forall j \in J \cup \{\zeta\}, \tag{5.1}$$

$$0 \leq \mu_{jl} \perp 1 - z_{jl} \geq 0, \forall j \in J \cup \{\zeta\}, \tag{5.2}$$

$$\lambda_l \perp \left( \sum_{j \in J \cup \{\zeta\}} z_{jl} - 1 \right) = 0, \tag{5.3}$$

$$0 \leq \rho_j^t \perp c_j - z_{jl} - \sum_{-l \in L_{od}^{t,n}, od \in OD, n \in \{1,2,\ldots,N\}} z_{j,-l} - f_j^{t-1} + g_j^t \geq 0, \forall j \in J, \tag{5.4}$$

where operator $\perp$ defines the complementarity relationship, i.e. $x \perp y \Leftrightarrow xy = 0$.

Since $\boldsymbol{\mu}_l = \mathbf{0}$ is always feasible to (5.1)-(5.4)[6], the above KKT conditions can be simplified as follows:

$$0 \leq z_{jl} \perp \phi_j - \lambda_l + \rho_j^t \geq 0, \forall j \in J \cup \{\zeta\}, \tag{6.1}$$

$$1 - z_{jl} \geq 0, \forall j \in J \cup \{\zeta\}, \tag{6.2}$$

and (5.3)-(5.4).

Note that at least one $z_{jl} > 0, \forall j \in J \cup \{\zeta\}$ due to (5.3), and it can be seen that any feasible solution to problem (3.1) and (3.3)-(3.5) has $\phi_{j^*} - \lambda_l + \rho_{j^*}^t = 0$, $\lambda_l = \phi_{j^*} + \rho_{j^*}$ and $z_{jl} = 0, \forall j \neq j^*$, where $j^* = \arg\min_{j \in J \cup \{\zeta\}} \{\phi_j + \rho_j^t\}$. Therefore $z_{j^*l} = 1$, $z_{jl} = 0$, $\forall j \neq j^*$. If we sum up (6.1) over set $L$ with the postulation that $\lambda_l = \lambda_{l'}$, we obtain

$$0 \leq \sum_{l \in L} z_{jl} \perp \sum_{l \in L} \phi_j + \sum_{l \in L} \rho_j^t - \sum_{l \in L} \lambda_l \geq 0, \forall j \in J \cup \{\zeta\},$$

and it further leads to

$$0 \leq h_j \perp \phi_j - u_L^t + \rho_j \geq 0, \forall j \in J \cup \{\zeta\}, \tag{7}$$

where $h_j = \sum_{l \in L} z_{jl}$ and $u_L^t = \lambda_l - \rho_j^t$.

As such, by combining the KKT conditions (for all users) and the market clearing conditions, it can be easily seen that (7) and (5.4) lead to the following lower level conditions (8.1)-(8.3) for all time intervals within each rolling horizon, which are essentially the lower level equilibrium constraints of the MPEC model. In addition, Constraints (8.4) present market clearing conditions.

*(lower level)*

$$0 \leq h_{j,od}^{t,n} \perp \left(np_j^t + \theta w_{jd} + \theta' v_{oj} + \rho_j^t\right) - u_{od}^{t,n} \geq 0, \ \forall j \in J, o \in O, d \in D, n \in \{1,2,\ldots N\}, t \in \{t_s, \ldots t_s + \tau\} \tag{8.1}$$

$$0 \leq h_{\zeta,od}^{t,n} \perp \lambda_{\zeta,od}^{t,n} - u_{od}^{t,n} \geq 0, \ \forall o \in O, d \in D, n \in \{1,2,\ldots N\}, t \in \{t_s, \ldots t_s + \tau\} \tag{8.2}$$

$$0 \leq c_j - f_j^t \perp \rho_j^t \geq 0, \ \forall j \in J, t \in \{t_s, \ldots t_s + \tau\} \tag{8.3}$$

$$0 \leq u_{od}^{t,n} \perp \sum_{j \in J \cup \{\zeta\}} h_{j,od}^{t,n} - H(u_{od}^{t,n}) \geq 0, \ \forall o \in O, d \in D, n \in \{1,2,\ldots N\}, t \in \{t_s, \ldots t_s + \tau\}. \tag{8.4}$$

Again, note that the above conditions (8.1)-(8.3) can also be postulated directly from Wardrop's principles, i.e. the users' behavior collectively satisfies Wardrop's first principle (Wardrop, 1952). Constraints (8.1) and (8.2) postulate user equilibrium, and Constraints (8.3) ensure users cannot violate the physical capacity of each parking area.

---

[6] Since it is $\mu_{jl} - \lambda_l, \forall j \in J$ that affect constraints (5.1) regardless of the individual values of $\boldsymbol{\mu}_l$ and $\boldsymbol{\lambda}_l$, $\boldsymbol{\mu}_l = \mathbf{0}$ is always feasible.

As a final remark, the results of the pricing model above include, for all time intervals, the price at each parking area, the occupancy level of each parking area, a record of when each user arrived and left their parking area, and the disutility of each type of user. These direct results can be used to derive metrics of the system's performance such as excess driving distance, the allocation of vehicles, the amount of demand that did not use the system (referred to as lost users), social surplus, and parking revenue.

### 3.3. Solution Method

The above dynamic MPEC model will be solved to obtain the optimal price of each parking area in each time interval. This difficult problem, given the dynamic time dimension, is approximated into multiple subproblems each involving only a subset of the time intervals. Such subproblems will be solved on a rolling horizon for all time intervals $t_s \in T$ while considering $\tau - 1$ intervals in the future. To develop the solution algorithm, we further assume that the inverse demand function $H(u) = a - b \cdot u$ is linear with respect to the equilibrium disutility $u$ for each type of demand. Parameters $a$ and $b$ are respectively the intercept of the demand curve and the demand elasticity.

For the model with only the occupancy objective (i.e. $\alpha_s, \alpha_r = 0$), the bi-level MPEC subproblems on a $\tau$ interval rolling horizon can be solved effectively by reformulation into an equivalent single-level mixed integer linear program (MILP) (for details, see Bai et al. 2012; 2014). Specifically, this method uses disjunctive integer constraints to reformulate the complementarity relationships.

More generally, for models with additional revenue or social surplus objectives, the bi-level MPEC subproblems involve non-convex bilinear revenue terms in the objective function (1.1), i.e. $\sum_{n \in \{1,2,...,N\}} \sum_{j \in J \cup \{\zeta\}} p_j^t \cdot n \cdot q_j^{t,n}$, which makes it highly difficult to find global optimal solutions. However, if the rolling horizon interval $\tau = 1$ is chosen (i.e. the model is solved interval-by-interval in a myopic way), this bilinear term can be reformulated into an equivalent series of linear and quadratic terms. This derivation, based on Hobbs et al. (2000), is presented as follows.

When $\tau = 1$, the subproblems are solved interval by interval, so $t = t_s$ in all subproblems $\forall t_s \in \{1,...,T\}$. So we further omit $\tau$ and $t_s$ in the derivation below for ease of notation.

First, the bilinear term can be rewritten as

$$\sum_{n \in \{1,2,...,N\}} \sum_{J \cup \{\zeta\}} p_j^t \cdot n \cdot q_j^{t,n} = \sum_{n \in \{1,2,...,N\}} \sum_{J \cup \{\zeta\}} n \cdot p_j^t \sum_{od \in OD} h_{j,od}^{t,n},$$

due to Constraints (1.5). If we sum up (8.1) over $n \in \{1,2,...,N\}, j \in J \cup \{\zeta\}, od \in OD$, we obtain the following equation:

$$\sum_{n \in \{1,2,...,N\}} \sum_{j \in J \cup \{\zeta\}} \sum_{od \in OD} h_{j,od}^{t,n} \cdot n \cdot p_j^t = \sum_{n \in \{1,2,...,N\}} \sum_{j \in J \cup \{\zeta\}} \sum_{od \in OD} h_{j,od}^{t,n} \cdot u_{od}^{t,n}$$

$$- \sum_{j \in J \cup \{\zeta\}} \sum_{n \in \{1,2,...,N\}} \sum_{od \in OD} h_{j,od}^{t,n} \cdot \left(\theta w_{jd} + \theta' v_{oj} + \rho_j^t\right).$$

Then based on (8.4), we have $\sum_{n \in \{1,2,...,N\}} \sum_{od \in OD} \sum_{j \in J \cup \{\zeta\}} h_{j,od}^{t,n} \cdot u_{od}^{t,n} = \sum_{n \in \{1,2,...,N\}} \sum_{od \in OD} H\left(u_{od}^{t,n}\right) \cdot u_{od}^{t,n}$. From (8.1)-(8.3) and (1.4), we further obtain

$$\sum_{j \in J \cup \{\zeta\}} \sum_{n \in \{1,2,...,N\}} \sum_{od \in OD} h_{j,od}^{t,n} \cdot \rho_j^t = \sum_{j \in J \cup \{\zeta\}} \rho_j^t \left( c_j - f_j^{t-1} + \underbrace{\sum_{m=\max(1,t-N)}^{t-1} q_j^{m,t-m}}_{g_j^t} \right).$$ Therefore, we are able to reformulate $\sum_{n \in \{1,2,...,N\}} \sum_{J \cup \{\zeta\}} p_j^t \cdot n \cdot q_j^{t,n}$ into the following linear and convex quadratic formulations:

$$\sum_{n \in \{1,2,...,N\}} \sum_{j \in J \cup \{\zeta\}} n \cdot p_j^t \cdot q_j^{t,n} = \sum_{n \in \{1,2,...,N\}} \sum_{od \in OD} H\left(u_{od}^{t,n}\right) \cdot u_{od}^{t,n} - \sum_{n \in \{1,2,...,N\}} \sum_{od \in OD} \sum_{j \in J \cup \{\zeta\}} h_{j,od}^{t,n} \cdot \left(\theta w_{jd} + \theta' v_{oj}\right)$$

$$- \sum_{n \in \{1,2,...,N\}} \sum_{od \in OD} \sum_{j \in J \cup \{\zeta\}} \rho_j^t \left( c_j - f_j^{t-1} + \underbrace{\sum_{m=\max(1,t-N)}^{t-1} q_j^{m,t-m}}_{g_j^t} \right).$$

With the assumption that $H\left(u_{od}^{t,n}\right)$ is a linear non-increasing function of $u$, the bi-level MPEC subproblems with the revenue objective for each single interval can be solved as an equivalent single-level convex mixed-integer quadratic program (MIQP). The reformulated MIQP is thus solvable through commercial solvers, albeit in a myopic fashion as opposed to over a rolling horizon. This approach does not apply when $\tau > 1$ due to the difficulty in formulating the future new users, $\sum_{n \in \{1,2,...,N\}} q_j^{t_s+1,n}$, in terms of the current occupancy and users leaving such that the MIQP is convex. We leave this challenging problem of $\tau > 1$ for future analysis.

## 4. Numerical Example

This section presents numerical examples to illustrate the improvements that the proposed dynamic pricing model could achieve as compared to current smart parking methods. We consider a test case based off the Marina neighborhood of the SF*park* pilot program as shown in Figure 1. The network has $|J| = 20$ parking areas including 282 parking spaces on 19 on-street block faces (blue lines) and 205 spaces in 1 large parking garage (represented as the filled-in blue P parking symbol). To simplify the model, rather than considering every possible origin (e.g. all TAZ centroids), the trip origins are aggregated at entry/access points to the parking neighborhood, by assuming that users have already made their mode choice and are only sensitive to local driving time within the neighborhood. The set of destinations would theoretically include all locations a user may want to go to, but for simplicity, they are aggregated at the city block level. In this example, there are three main travel destinations (shown as the green stars) and two travel origins, one at each of the two major access points (illustrated by the red cars). The test case is run over the entire enforceable day (9:00am – 6:00pm) with parking prices updated every 15 minutes.

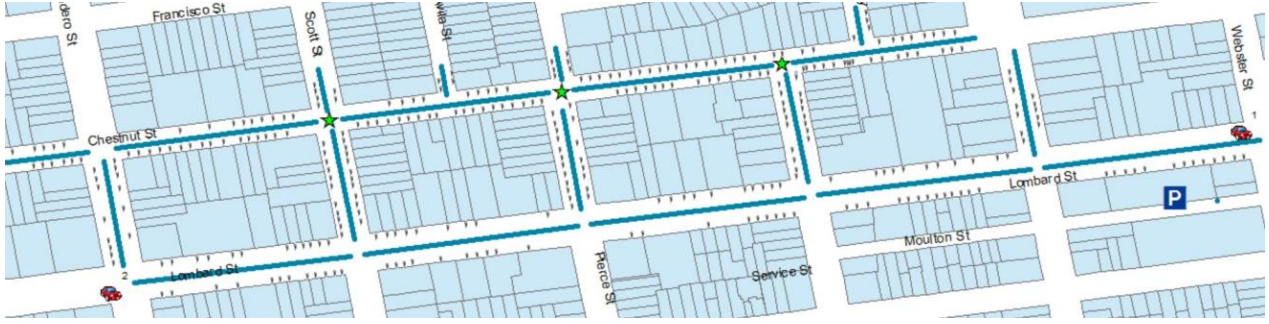
Figure 1 - Study area (Marina neighborhood, San Francisco, CA)

The value of walking (access) time was taken to be $\theta = \$0.42$/minute, which is 80% of the average hourly wage (\$31.77) in the San Francisco area (Bureau of Labor Statistics, 2013), and the value of driving time was taken to be $\theta' = \$0.26$/minute, or 50% of the average hourly wage (O'Sullivan, 2012). It is assumed that the parking price can only increase or decrease at most \$1/hour between each time interval (i.e. $\varepsilon_r = \varepsilon_l = \$1$/hour). However, no minimum or maximum price limit was enforced in this study (i.e. $\gamma = 0$ and $\eta = \infty$).

The demand data over time are estimated from TAZ-level O-D estimates provided by SFCTA (2009) from their activity-based travel demand model SF-CHAMP. Vehicle trip data for the five different times-of-day periods, i.e. $t_p$ = early AM, AM peak, midday, PM peak, evening, as defined by SFCTA (2009), were used to fit a multimodal distribution (9) as an approximation of the time-dependent user demand. The elasticity coefficient of the linear demand function $H(\cdot)$ is assumed to be $b = -0.3$ (TCRP, 2005), which is very close to Price and Shoup's (2013) value of $-0.28$ for the Marina neighborhood. The linear demand curve's intercept $a_{od}^{t,n}$, for all $t$, $n$, and $od$, is assumed to be a function of several properties of typical time-dependent demand; i.e.,

$$a_{od}^{t,n} = k\left[ m_{od} \sum_{t_p=1}^{5} \frac{P_{t_p}}{\sqrt{\sigma_{t_p}}} e^{\left(-\frac{t-\mu_{t_p}}{\sigma_{t_p}}\right)^2} + N(0, 0.25) \right] v_n, \qquad (9)$$

where $k$ is the estimated percent of total vehicle trips to the parking neighborhood that use the proposed parking management system, $m_{od}$ is the fraction of the vehicle trips that have *OD* pair *od*, $P_{tp}$ is the total number of trips to the parking neighborhood during time-of-day period $t_p$, $\sigma_t$ is the length of the time-of-day period in hours, $\mu_t$ is the middle hour of the time-of-day period, and $v_n$ is the fraction of trips that have parking duration $n$. To determine $m_{od}$, it is assumed that two thirds of the users enter the study area along the main arterial from the east, with the rest entering from the west. The destinations are each weighted similarly, with 40% attracted to the middle destination and 30% to each of the two outer destinations. The desired parking duration can take one of five discrete values: 15 minutes, 45 minutes, 1.5 hours, 3 hours, or 4 hours. The distribution of the parking duration was borrowed from a study of urban parking in Sofia, Bulgaria (Naydenov, 2010). Normally distributed random errors were added to each O/D demand and then rounded to the nearest whole number; see (9). Using the resulting values of $a_{od}^{t,n}$ as a medium benchmark demand level, we also consider a low demand case and a high demand case which uniformly scale the benchmark demand (9) by 0.5 and 2, respectively, before introducing the randomness term.

Four different parking management scenarios are studied: (i) the *traditional* scenario where drivers have no information at all; (ii) the *static information* scenario where drivers check parking availability and pricing information only when they start their trips; (iii) the *dynamic information* scenario where drivers are informed with updated parking space availability while driving; and (iv) the *dynamic pricing* scenario where drivers can check updated near real-time availability information and prices are allowed to change at every time interval. Scenarios (i)-(iii) all involve static parking prices. Following SF*park*, a typical day was split into three periods, morning (9am-12pm), early afternoon (12pm-3pm), and early evening (3pm-6pm). The static parking price in each period for scenarios (i)-(iii) was set to be equal to the average price over the corresponding time intervals from scenario (iv), so as to have fair performance comparison. Scenarios (i)-(iii) were simulated in MATLAB using the logic similar to the agent-based PARKAGENT model (see Levy et al., 2012). Users choose parking areas according to the same linear demand function as the dynamic pricing scenario but may not have complete information in terms of location-specific price or current availability depending on the scenario. After users fail to park in a fully occupied parking area, they "cruise" to the next preferred parking area until they give up after having tried 75% of the areas in the entire neighborhood.[7]

The proposed dynamic pricing model in scenario (iv) was solved on a rolling horizon ($\tau=2$) (when possible) and with three different objectives: (a) the occupancy objective ($\alpha_s, \alpha_r=0, \beta=2.5$, and $\tau=2$) which pushes occupancy levels as close to 85% as possible everywhere; (b) the social surplus objective ($\alpha_s=1, \alpha_r=0, \beta=2.5$, and $\tau=2$) which balances the occupancy objective with the additional goal of maximizing the system's social surplus; and (3) the revenue objective ($\alpha_r=1, \alpha_s=0, \beta=2.5$, and $\tau=1$) which balances the occupancy objective with the additional goal of maximizing revenue. The reformulated equivalent MIQP of the dynamic pricing model (1.1)-(1.6), (8.1)-(8.4) was solved by the commercial solvers, Gurobi and CPLEX, on a personal computer with 3.1GHz CPU and 16GB RAM.

Table 1 - Numerical results under three different levels of demand.

| Demand Level | Performance Metric | Scenarios | | | | | |
|---|---|---|---|---|---|---|---|
| | | Static Pricing | | | Dynamic Pricing (iv) | | |
| | | (i) Traditional | (ii) Static Information | (iii) Dynamic Information | (a) Occupancy | (b) Social Surplus | (c) Revenue |
| Low | Total excess distance *(miles)* | 122.64 | 48.52 | 37.33 | 0.00 | 0.65 | 0.00 |
| | Occupancy distribution *% lot-hrs empty/above target* | 39.9/48.1 | 26.1/32.9 | 26.7/31.9 | 26.9/0.0 | 8.1/0.6 | 15.7/0.0 |
| | Lost customers | 197 | 197 | 197 | 197 | 224 | 418 |
| | Social surplus *(utils)* | 3,045 | 3,242 | 3,251 | 3,337 | 3,413 | 3,029 |
| | Parking revenue | $251.97 | $190.71 | $189.49 | $129.23 | $190.80 | $601.24 |
| Medium | Total excess distance | 522.27 | 257.99 | 166.13 | 0.00 | 5.54 | 0.92 |
| | Occupancy distribution | 20.6/64.0 | 7.2/51.7 | 8.2/50.0 | 6.3/0.0 | 7.1/1.7 | 21.4/0.3 |
| | Lost customers | 460 | 414 | 402 | 402 | 410 | 767 |
| | Social surplus | 8,946 | 9,401 | 9,414 | 9,724 | 10,056 | 8,824 |
| | Parking revenue | $1,050.38 | $970.33 | $970.33 | $611.07 | $661.49 | $1,840.78 |
| High | Total excess distance | 1,622.23 | 1,213.13 | 721.22 | 4.91 | 346.48 | 3.69 |
| | Occupancy distribution | 5.4/79.0 | 0.7/80.7 | 0.7/81.5 | 1.8/0.6 | 2.1/39.2 | 12.8/0.4 |
| | Lost customers | 1,514 | 1,150 | 1,121 | 954 | 927 | 1,571 |
| | Social surplus | 25,774 | 26,508 | 26,404 | 27,946 | 28,971 | 24,896 |
| | Parking revenue | $4,033.71 | $4,765.75 | $4,847.42 | $3,637.19 | $3,249.96 | $5,948.67 |

---

[7] A higher "willingness-to-try" percentage decreases the number of users lost, but may increase the total excess distance driven.

The computational results are summarized in Table 1. The proposed dynamic pricing model is found to generate a number of significant benefits in terms of a reduction in excess driving distance, a more balanced occupancy distribution, less lost parking demand due to users' inability to find open parking spaces, and increased social surplus (which can be balanced between consumer and producer surplus, i.e. revenue). We provide a more detailed discussion below.

**Circling and congestion reduction.** Recall that the primary goal of adjusting parking prices in real-time is to eliminate circling -- by ensuring parking spaces are always available at each parking area. This is exactly the objective of dynamic pricing model iv-(a), and Table 1 shows significant reductions in excess driving distance. At low demand levels, all users are able to find a space that suits their preferences on the first try, such that there is strictly no excess driving distance. Adding alternative terms to the objective (such as scenario iv-(b), and iv-(c)) puts more weight on conflicting goals and can result in some circling. A high weight on social surplus, as in model iv-(b), tends to seek small gains in total social surplus at the expense of the occupancy objective and congestion reduction. In model iv-(c), there is less excess driving distance within the neighborhood, but due to the significantly higher number of lost users, this congestion is likely being diverted to adjacent parking areas (which is not captured in our simulation) and may actually be worse than model iv-(a).

In terms of excess driving distance, the static and dynamic information models show rather large improvements over the traditional parking model. This is expected and is consistent with established research. For example, based off Shoup's (2006) estimate for the excess VMT generated per space per year due to circling, this one-day simulation would generate 3,807 VMT from the 282 on-street parking spaces, plus up to 2,767 VMT from the 205 off-street spaces. Because the prices for the static pricing are set to equal the average prices from the dynamic pricing model which is designed for reducing congestion and emissions, we expect the excess distance values of the simulated scenarios (i)-(iii) (although on similar scales) to be less than these estimates, as is shown in our results.

Our results (for scenario iii) are also consistent with SF*park*'s evaluation (Millard-Ball et al., 2014) which estimates that SF*park* reduced circling (or excess driving distance) by about 50% since its inception (our results suggest 70%, 68%, and 56% for low, medium, and high demand levels, respectively). Our evaluation also supports that 85% occupancy is an effective threshold for reducing circling.

**Parking space occupancy.** The dynamic pricing model is also able to use prices as a leverage to influence users to change parking location based on real-time demand patterns. This allocation of parking leaves fewer lot-hours with zero occupancy (i.e. empty), and almost no lot-hours exceeding the target occupancy level in low to medium demand cases. The ideal parking allocation would maintain an occupancy level very close to the target level throughout the system at all times despite very high or very low demand. The dynamic pricing model was found to be the most capable of achieving this ideal, as indicated by the high amount of lot-hours that are not empty and yet still below the occupancy target.

**Lost users.** As discussed in Section 3.2, some users will inevitably be "lost" to parking areas outside of the system (e.g. private garages, hotels, restaurants). When comparing different parking management strategies, if they all have the same average price they should all have about the same number of users lost due to the elastic demand curve. However, users can also be lost when they cannot find an available parking space and give up looking, either taking their business elsewhere or going home, which is undesirable. So we focus on measuring the number of lost users due to capacity limitation, i.e. after they have tried 75% of the lots in the neighborhood and give up. Although the number presented in Table 1 represents both types of lost users, the differences in this number among the four scenarios are mainly due to capacity limitation.

From the numerical results, the number of lost users is comparable in all scenarios. In high-demand static-price cases, a higher number of users reach the limit of the number of lots they will try. Dynamic pricing, on the other hand, is able to allocate users more effectively to minimize the number of full lots, and in turn the number of lost users. We believe the improved allocation of parking helps retain more users and could benefit the local economy.

**Social surplus.** Another possible metric to consider is the total social surplus. This is maximized when the price is zero, which conflicts with the occupancy objective. Therefore, a large weight on social surplus, as in model iv-(b), could cause poor parking allocations and worsen congestion. Model iv-(a) already tends to produce more social surplus than the static price scenarios, so unless emissions or external economic benefits were incorporated into the definition of social surplus (they are not in these results), then it is better to implement the iv-(a) or iv-(c) objectives.

**Revenue.** Because the occupancy objective already helps improve social surplus without violating the goals of congestion mitigation, a more reasonable secondary objective is to maximize revenue. Like the social surplus objective, this objective also conflicts with the occupancy objective, but in the opposite direction. Maximum revenue is achieved by raising prices, thus losing users to elastic demand and deviating from the occupancy objective towards an underutilized allocation, until the marginal revenue equals zero (because the supply curve is constant). This results in occupancy levels near 50% (under the current linear demand settings) in our numerical examples. Despite the low occupancy level (as seen in the high demand case in Table 1), much more (e.g. up to 65%) users are lost. Therefore, the revenue objective should have a small weight as long as any revenue targets (if any) could be met. As mentioned, a weight of $\alpha_r=1$ was used for the results in Table 1. For example, for the high demand case, if the weight is changed to $\alpha_r=1/3$, the lost users can be decreased 18% while the revenue only decreases 8% and the parking allocation improves slightly. In the low and medium demand cases, a weight of $\alpha_r=1/3$ decreases both the revenue and the lost users about equally (22% and 16%, respectively). The revenue objective is more useful for high demand cases where price plays a large role in determining the parking allocation. In low demand cases where many lots will be free anyway, the occupancy objective produces more evenly distributed parking allocations.

## 5. Conclusion

The dynamic pricing model developed in this paper shows promise to reduce vehicle circling and congestion in busy urban centers and positively impact the local economy. By utilizing real-time demand information, prices can be updated and disseminated periodically throughout the day to influence drivers' parking decisions and better allocate vehicles. By maintaining an occupancy level below 85% on each block, circling for parking is found to be practically eliminated. This may not have been possible when pricing tables are only updated every six weeks (as does the most advanced parking system to date).

The proposed dynamic Stackelberg leader-follower game pricing model can effectively capture multiple drivers' individual choices while making pricing decisions to achieve beneficial system objectives. Our numerical examples show that this modelling framework can balance parking occupancy, reduce excess driving distances, lower the number of frustrated users, improve social surplus, and potentially increase parking revenue.

One significant limitation of the proposed model is the omission of competition with other parking agencies, i.e. "outside lots." As we mentioned in the paper, if the locations, prices and capacities of the outside lots are fixed and given, it is trivial to incorporate them in the model. However, if there are multiple agencies making pricing decisions at the same time, considering them on top of the already highly complex bi-level problem would be a major challenge. Nevertheless, the proposed model provides a useful building block for addressing such situations in the future. Future efforts may also consider multiple vehicle types, such as motorcycles, compact or electric vehicles, commercial vehicles, etc., or user types, such as shoppers, tourists, and commuters who have different price elasticities or even demand curves. Additionally, the model may consider the location of traffic congestion (e.g. those generated by vehicle circling) and its impacts on the non-parking traffic. Before this model can be used by a parking agency, the parameters in the model (e.g. those related to drivers' travel and parking behavior) must be calibrated with empirical data. This may have to be done for each parking agency, but smaller agencies could benefit from initial results from cities such as San Francisco, Los Angeles, or Chicago who have already been successful in implementing innovative parking management ideas in recent years. The model's accuracy can be further improved by using a more advanced demand predication model based on both historic data and real-time information collected through sensors and/or end-user devices. Ultimately, we hope this model brings significant benefits in terms of reduction of vehicular emissions, growth in the urban economy, improvements in transportation mobility, and progress towards sustainable and equitable transportation policies.

## Acknowledgments
The authors would like to thank Kevin Fuller (University of Illinois) for help with a preliminary literature review on urban parking systems, and for support throughout the project. The authors would also like to thank SFCTA for providing data for the case study, and SF*park* for making their data publicly available. This research was supported in part by the U.S. National Science Foundation via Grant CMMI-0748067 and the Department of Civil and Environmental Engineering at the University of Illinois at Urbana-Champaign via a departmental Innovation Research Grant.

3. Ayala, D., Wolfson, O., Xu, B., Dasgupta, B., and Lin, J. (2012). Pricing of parking for congestion reduction. Proceedings of the 20th ACM SIGSPATIAL International Conference on Advances in Geographic Information Systems (ACM GIS) Redondo Beach, CA, November 2012.

4. Bai, L., Mitchell, J.E. and Pang, J.S. (2013). On convex quadratic programs with linear complementarity constraints. *Computational Optimization and Applications,* 54(3), 517-554.

5. Bai, Y., Ouyang, Y., and Pang, J.S. (2012). Biofuel supply chain design under competitive agricultural land use and feedstock market equilibrium. *Energy Economics*, 34(5), 1623–1633.

6. Bai, Y., Ouyang, Y., and Pang, J.S. (2014). Enhanced models and improved solution for competitive biofuel supply chain design under land use constraints. *European Journal of Operational Research.* Under review.

7. Bureau of Labor Statistics (2013). Occupational employment and wages in San Francisco-San Mateo-Redwood City, May 2012. 13-811-SAN. Available at: http://www.bls.gov/ro9/oessanf.htm. Accessed May 6, 2014.

8. Caicedo, F., Blazquez, C., and Miranda, P. (2012). Prediction of parking space availability in real time. *Expert Systems with Applications*, 39(8), 7281-7290.

9. Daniel, J.I. (1995). Congestion pricing and capacity of large hub airports: A bottleneck model with stochastic queues. *Econometrica: Journal of the Econometric Society*, 327-370.

10. Facchinei, F. and Pang, J.S. (2003). *Finite-Dimensional Variational Inequalities and Complementarity Problems*, Volumes I and II. Springer-Verlag, New York.

11. Geng, Y. and Cassandras, C.G. (2011). Dynamic resource allocation in urban settings: a "smart parking" approach. Proceedings of 2011 IEEE Multi-Conference on Systems and Control, October 2011.

12. Giuliano, G. (1994). Equity and fairness considerations of congestion pricing. Report 242. Transportation Research Board.

13. Hobbs, B., Metzler, C., and Pang, J.S. (2000). Strategic gaming analysis for electric power networks: an MPEC approach. *IEEE Transportation Power Systems*, 15(2), 638–645.

14. Hu, J., Mitchell, J.E., Pang, J.S., Bennett, K. and Kunapuli, G. (2008). On the global solution of linear programs with linear complementarity constraints. *SIAM Journal on Optimization*, 19(1), 445–471, 2008.

15. Hu, J., Mitchell, J.E., Pang, J.S. and Yu, B. (2012). On linear programs with linear complementarity constraints. *Journal on Global Optimization*, 53(1), 29-51.

16. IBM (2011). Global Parking Survey. Available at: http://www-03.ibm.com/press/us/en/pressrelease/35515.wss. Accessed on November 17, 2013.

17. Intergovernmental Panel on Climate Change (2007) "Climate change 2007: Impacts, adaptation, and vulnerability", Fourth assessment report, April 2007. Available at: http://www.ipcc.ch. Accessed November 17, 2013.

18. Levy, N., Martens, K., and Benenson, I. (2012). Exploring cruising using agent-based and analytical models of parking. *Transportmetrica A: Transport Science,* 9(9), 773-797.
17

19. Litman, T. (2005). Transportation cost and benefit analysis guidebook: Techniques, estimates and implications. *VTPI*.

20. Liu, W., Yang, H., and Yin, Y. (2013). Expirable parking reservations for managing morning commute with bottleneck congestion and limited parking spaces. Working paper.

21. Lu, Y. (2006). Robust transportation network design under user equilibrium. Massachusetts Institute of Technology. Master's thesis.

22. Luo, Z.Q., Pang, J.S., and Ralph, D. (1996). Mathematical Programs with Equilibrium Constraints. Cambridge University Press, Cambridge, England.

23. Mackowski, D., Bai, Y., Ouyang, Y., and Fuller, K. (2014). Informatics-enabled urban management: Envisioning the next generation of "smart parking." Working Paper. University of Illinois at Urbana-Champaign.

24. Millard-Ball, A., Weinberger, R.R., and Hampshire, R.C. (2014). Is the curb 80% full or 20% empty? Assessing the impacts of San Francisco's parking pricing experiment. *Transportation Research Part A: Policy and Practice*, 63(2014), 76-92.

25. Naydenov, N. (2010). Parking survey – methodology and case study. FIG Congress 2010 Sydney, Australia, April 2010.

26. O'Sullivan, A. (2012). Urban Transit. *Urban Economics* (8 ed., pp. 290-314). New York: McGraw-Hill.

27. Pierce, G and Shoup, D. (2013). Getting the prices right. *Journal of the American Planning Association*, 79(1), 67-81.

28. Qian, Z. and Rajagopala, R. (2013). Optimal parking pricing in general networks with provision of occupancy information. *Procedia-Social and Behavioral Sciences*, 80, 1–26.

29. Qian, Z. and Rajagopala, R. (2014). Optimal occupancy-driven parking pricing under demand uncertainties and traveler heterogeneity: A stochastic control approach. *Transportation Research Part B: Methodological*, 67(2014), 144-165.

30. San Francisco County Transportation Authority (2009). SF-CHAMP Base OD. Available at: http://www.sfcta.org/downloads/serviceBureau/. Accessed on January 13, 2014.

31. Shoup, D. (2005). The high cost of free parking. *American Planning Association*: Chicago, IL.

32. Shoup, D. (2006). Cruising for parking. *Transport Policy*, 13(6), 479-486.

33. Small, K.A. (1992). Using the revenues from congestion pricing. *Transportation*, 19(4), 359-381.

34. Stefik, M.J., Bell, A.G., Eldershaw, C., Good, L.E., Greene, D.H., Torres, F.E., Uckun, S., and Cummins, D.P. (2014). *U.S. Patent No. 8,671,002*. Washington, DC: U.S. Patent and Trademark Office.

35. TCRP (2005). *Traveler Response to Transportation System Changes Handbook*, Third Edition: Chapter 13, Parking Pricing and Fees, Report 95. Transportation Research Board.

36. Yang, H. and Bell, M.G. (2001). Transport bilevel programming problems: recent methodological advances. *Transportation Research Part B: Methodological*, 35(1), 1-4.